# SANet:Superpixel Attention Network for Skin Lesion Attributes Detection

Paper ID

**Abstract**

The accurate detection of lesion attributes is meaningful for both the computer-aid diagnosis system and dermatologists' decisions. However, unlike lesion segmentation and melanoma classification, there are few deep learning methods and literatures focusing on this task. Currently, the lesion attribute detection still remains challenging due to the extremely unbalanced class distribution and insufficient samples, as well as large intra-class and low inter-class variations. To solve these problems, we propose a deep learning framework named superpixel attention network (SANet). Firstly, we segment input images into small regions and shuffle the obtained regions by the random shuttle mechanism (RSM). Secondly, we apply the SANet to capture discriminative features and reconstruct input images. Specifically, SANet contains two sub-modules: superpixel average pooling and superpixel attention module. We introduce a superpixel average pooling to reformulate the superpixel classification problem as a superpixel segmentation problem and a SAMis utilized to focus on discriminative superpixel regions and feature channels. Finally, we design a novel but effective loss, namely global balancing loss to address the serious data imbalance in ISIC 2018 Task 2 lesion attributes detection dataset. The proposed method achieves quite good performance on the ISIC 2018 Task 2 challenge.

## Introduction

Melanoma has attracted extensive attentions during recent decades, since it is one of the most rapidly rising and deadliest skin cancers. To distinguish melanoma, dermatologists developed several methods to aid diagnosing skin lesion, such as 'ABCD rules', 'Menzies method' and '7-point checklist'. In general, all these methods highly depended on the detections of attributes of skin lesions (Oliveira et al. 2016; Nachbar et al. 1994). The attribute detection made by human visual inspection is often difficult, time-consuming, and subjective, which requires computer-aided diagnosis (CAD) system for the attributes detection. In addition, CAD system also needs the detection of attributes. According to the recent research, introducing lesion attributes into CAD systems can significantly improve the performance (Jahanifar. et al. 2018). Hence, to increase both the efficiency and accuracy of dermatologists' diagnosis, the automated detections of skin lesion attributes are highly desired.

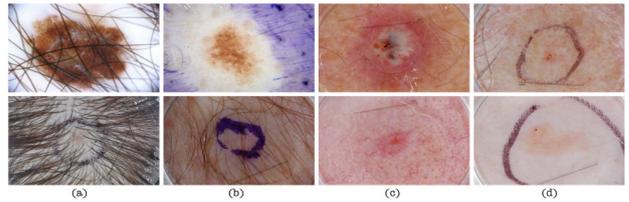

Figure 1. Challenges for automated detection of lesion attributes.

Lesion attributes (a.k.a., dermoscopic structure), are meaningful visual skin lesion pattern in lesion texture (Lopez-Labraca et al. 2018). Each of them is varied based on the lesion type and malignancy severity. In ISIC 2018 dataset, five of the most clinically interesting lesion attributes, pigment networks, globules, milia like cysts, negative networks, and streaks, are labeled for training. Some of them may appear on only one superpixel (Achanta et al. 2012). Some may occupy nearly whole lesion area. Moreover, only the attributes within the most salient lesion are labeled.

The automated detections for ISIC 2018 skin lesion attributes remain challenging due to several reasons. First, as shown in Figure 1, the existence of occlusion including: (a) thick hair, (b) dye concentration, (c) bloods vessels and (d) clinic markers can easily cause the indistinction problems. Second, it is difficult to build an end-to-end neural network to deal with superpixel classification problems and capture attributes. Finally, training samples are extremely unbalanced and scared, which further prevent models achieving better generalization capabilities.

To address above restrictions, we propose SANet which involves three components: (1) region shuttle mechanism (RSM); (2) superpixel attention module (SAM); (3) global balancing loss (GBL). The RSM are designed to augment training date and learn discriminative local visual features. For example, milia like cyst usually occupies several superpixels. To distinct milia like cyst requires networks in the human style, the networks are required to acquire informative local patterns. SANet tries to recover the deconstructed one and focus on informative regions by incorporating superpixel average pooling and superpixel attention module. GBL, is the loss combines global balancing cross entropy loss and global balancing jaccard loss. Optimizing the network with global balancing network can achieve higher jaccard index. For the reason that the global balancing network is with the properties of imbalance classes and foreground/background ratio:

(1) We propose a SANet, which incorporates superpixel average pooling and superpixel attention module. The proposed model can also reformulate super-pixel classification problems as the superpixel segmentations and superpixel attention mechanism is leveraged to enhance the discriminative of visual features.
(2) A novel global balancing loss is proposed to avoid the imbalance of training data and enable our network to focus on informative samples.

## Related Work

The development of UNet made breakthrough in medical image segmentation (Ronneberger et al. 2017). Afterward, researchers proposed a lot of improvements based on UNet. Most of them are divided into two categories: (1) more representative visual features are learned by adopting deeper backbones. (2) The attention mechanism is adopted to select more discriminative regions and features.

### UNet

UNet and the variants of the encoder-decoder architecture have achieved significant improvements on medical image segmentations through skip-connections and up-sampling layers (Long et al. 2015). With skip-connections and up-sampling layers, UNet can utilize early layers' information to refine high-level features and output prediction with the same resolution as inputs. Although the encoder-decoder architecture always obtains high accurate boundary segmentation performance, due to skip-connections and up-sampling operation, the results lack consistency with the operation of skip-connections and up-sampling. To remedy the above problems, we propose a method, called SANet.

### Backbone

Due to the success of deep neural network in ImageNet datasets, backbone of UNet shift from VGG 16 to 50 layers, 101 and 151 layers Residual Network (ResNet 50, 100, 151), 169 layers Densely connected Network (DenseNet 169) (He et al. 2018; Huang et al. 2016). ResUNet and H-DenseUNet were shown significant performance boosts in visual features learning. (Li et al. 2018; Jin et al. 2019; Zhang et al. 2018) utilized deformable convolutional network to extract context information (high level semantic features) and obtain precise localization (low level local features). (Guo et al. 2018) proposed stacked dense UNet aligning human faces with scale invariant features.

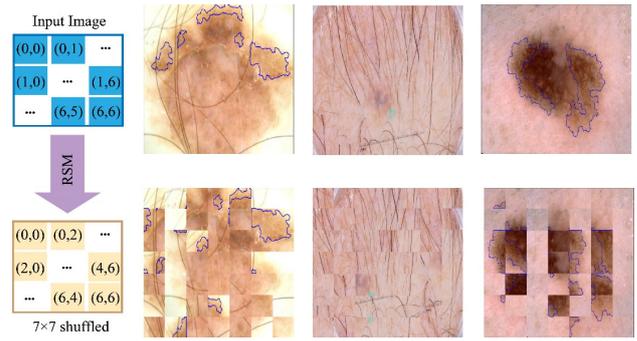

Figure 2: These are examples of skin lesion images (Top) and corresponding shuffled images (bottom). The purple lines show the labels of pigment networks and the cyan ones show that of milia like cysts.

### Random shuffle module

Local information of an input image is really helpful for medical image segmentation. Recently, (Chen et al. 2019) won the first place in fine grained images classification challenges: iMaterialist and FieldGuide by inputting deconstructed images. To learn local discriminative features, they divided an input image into several local region and assemble them randomly. However, as shown in Figure 2, RSM introduces noisy visual skin lesion patterns (the checkboard problem) from shuffled input images and noisy supervision signal from shuffled labels. In this work, we adopt superpixel average pooling and SAM jointly to alleviate such problems.

### Attention module

Inspired by the channel attention mechanism in squeeze network, a large amount of attention mechanisms has been introduced to select discriminative regions and feature. Abhijit *et al.* proposed concurrent spatial and channel 'Squeeze & Excitation' to recalibrate features from the channel and spatial dimension (Roy et al. 2018; Iandola et al. 2016). By learning where to look, (Oktay et al. 2018) proposed a model to depresses irrelevant regions for the given an input image and highlight salient features. In this paper, we introduce a novel attention mechanism named SAM to focus on significant superpixel region to boost segmentation performance.

### Loss for medical image segmentation

Apart from the improvements of structure for basic UNet, researchers also designed different losses for different tasks in the field of medical image analysis. Kawahara *et al.* proposed a novel multi-channel Dice loss (Sudre et al. 2017). The modified jaccard loss was first presented for ISIC skin lesion detection challenge (Yu et al. 2017). To remedy the imbalance between foreground pixels and background pixels, focal loss is proposed by Lin *et al.* down-weights easy example and focus on informative samples automatically (Lin et al. 2017).

## Methodology

In this section, we briefly introduce a random shuffle module. We then discuss how to embed the SAM into the encoder-decoder model. Finally, we present a novel loss function called the global balancing loss.

### Random shuffle module

Inspired by the nature language processing idea of learning local discriminative words by shuffling words in a sentence, the destruction and construction structure forces network to learn discriminative regions for fine grained classification tasks by shuffling image regions (Yu et al. 2018).

As shown in Figure 2, we employ random shuffle module to deconstruct the structure of skin lesions. First, we partition an input image into $7 \times 7$ sub-regions. Second, shuffling sub-regions with their nearest neighbors according to uniform distribution in a row-major order. Finally, these random shuffled sub-regions are assembled. Noted, in order to avoid that the original images are overwhelmed by the destructed images, we set the ratio of the destructed images in batch to $1:1$.

### Superpixel attention network

The superpixel attention network consists of Residual UNet (ResUNet), superpixel average pooling and superpixel attention module.

#### Residual UNet

Deep convolutional neural network (DCNN) is powerful in terms of medical image analysis. However, training DCNN from scratch is a challenging task due to various reasons, such as the robust design of network architecture, proper weight initialization, an optimum learning rate setting and weight decay. In this paper, we adopt the transfer learning to circumvent these obstacles. Our framework combines the ResNet 101 and UNet. The ResNet 101 (pre-trained on ImageNet dataset) is explored as the encoder for the proposed segmentation model to extract features. The outputs are obtained from the last convolutional layer of residual network, and then are fed to the superpixel attention module.

#### Superpixel Average Pooling

The task of detecting lesion attributes is to classify the category of the selected superpixel. Given a dermoscopy image, and the corresponding superpixel map $s$, the purpose is to classify each superpixel into five potentially overlapping lesion attributes: pigment network, negative network, milia like cysts, streaks and globules.

If we want to reuse the previous methods, such as UNet and Deeplab (Chen et al. 2018), we need to transform this proposed network to complete a segmentation task. To convert the superpixel classification task to the superpixel segmentation task, we design a novel pooling method called superpixel average pooling, which enables superpixel segmentation models to train via an end-to-end way. Given the feature map $F^l$, we first aggregate the values within the $i$-th superpixel $s_i$, and then we divide them by the number of pixels within the region $s_i$, yielding a superpixel regional feature $R_i$:

$$R_i \frac{\sum_{(h,w) \in S_i} F^l(h,w)}{\sum_{(h,w) \in S_i} 1} \qquad (1)$$

### Superpixels attention module

Lesion attributes have low inter-class variations and huge intra-class differences. Lesion attributes of the same types may vary hugely according to patients' ages and ethics, lesion positions, lighting, scales, devices' types and views (Zhen et al.). For ISIC 2018 skin lesions challenge, they are acquired from all anatomic sites with a variety of dermascopic types, historical samples of patients and several different institutions. These differences lead to numerous variations, which hamper neural network to detect lesion attributes accurately. Moreover, different attributes follow divergent grow patterns. For example, pigment network tends to indicate benign lesions and the negative network commonly represents melanoma. The milia like cyst may scatter the whole lesion area. As for globules and streaks, they always grow up from the inner boundary of the lesion.

To embed complex boundary and contextual information, we explore a novel SAM for medical image segmentation. Noted that we append SAM after the superpixel average pooling. Figure 3(c) illustrates the structure of the proposed SAM. We first utilize a convolutional layer to handle the features extracted by the ResUNet. Then we feed the features into the SAM to generate new features with both boundary and contextual information through the two following steps. The first step is to generate superpixel attention features by unpooling superpixel regional features. Spatial attention matrix is generated by performing attention operation. For attention operation, we first perform concatenation between the original input features and the superpixel attention features. Second, we perform a matrix convolution over the concatenated features to gain superpixel attention features.

$$\mathbf{P} = w * (Unpooling(\mathbf{R}) + \mathbf{F}), \qquad (1)$$

where $\mathbf{R} \in \mathbb{R}^{C \times W}$ is superpixel regional features, $\mathbf{F} \in \mathbb{R}^{C \times H \times W}$ is the input features and $C$ is the number of feature channel, $H$ and $W$ are the height and weight of feature $N$ is the number of superpixels. $\mathbf{P}$ is superpixel attention features of the final output SAM, $w$ is the parameter used to recalibrate the superpixel regional features and the input features.

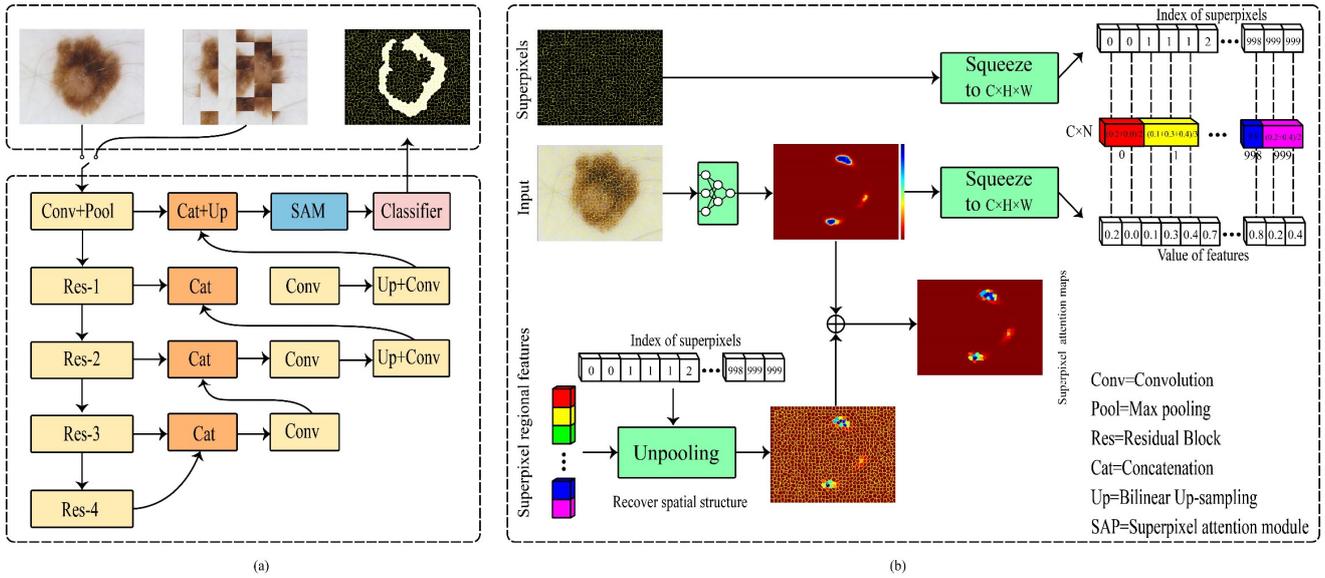

Figure 2. (a) Proposed SANet; (b) Superpixel average pooling; and (c) Superpixel attention module.

**Global balancing loss**

The classic segmentation losses tend to be less accurate when the training dataset is imbalanced, such as cross entropy loss and dice loss. As shown in Table I, there is a huge imbalance problem for the detection task of the attributes. First, the attributes face a serious intra-class imbalance, where the ratio between foreground and background are high. Second, there is a serious imbalance among classes. The imbalance causes problems to these losses including: (1) it is easy to occur over-fitting when the ratio between foreground and background are imbalanced for cross entropy, (2) the background pixels' signals are easily ignored for jaccard loss, which can lead to under-segmentation during the inference stage. Besides, the jaccard loss of multi-class version is calculated by the average jaccard loss over different channels, which causes the network more sensitive to outliers lying in classes with relative small number of the foreground pixels.

To overcome the drawbacks of cross entropy loss and dice loss, researchers combined them together and achieved better accuracy in many segmentation tasks, including 2018 Data Science Bowl, the task I lesion segmentation of ISIC 2018 skin lesion detection.

$$L_{all} = 0.5 * CEL + 0.5 * JAL \quad (2)$$

where $CEL$ is cross entropy loss and $JAL$ is jaccard loss.

**Global balancing Jaccard loss**

Jaccard index ordinary is a measure of assessing the overlap between prediction and ground truth. Milletari *et al.* proposed a dice loss as a loss function, and in ISIC 2016 challenge people proposes a jaccard loss based on a dice loss and jaccard index.

$$JAL = 1 - \frac{\sum_{i=1}^{n} p_i q_i}{\sum_{i=1}^{n} p_i^2 + \sum_{i=1}^{n} q_i^2} \quad (3)$$

where $p_i$ denotes the predicted probability of $i-th$ superpixel and $q_i$ denotes the corresponding ground truth.

However, JAL is used to solve one class segmentation (i.e., segmenting foreground and background). Later, the multi-class dice loss and ja loss was developed. However, multi-class jaccard loss is just the average version of jaccard loss. When face blanked images, multi class jarrcard loss could face divergent. In this paper, we propose a global balancing jaccard loss, which is defined as:

$$GDJAL = 0.5 * JAL_{micro} + 0.5 * JAL_{macro} \quad (4)$$

where $JAL_{macro}$ is macro jaccard loss, which is a popular multi-class jaccard loss. It conducts jaccard loss independently for each class and then calculate the average among them. $JAL_{micro}$ is micro jaccard loss, which aggregates positive pixels of all classes to calculate the final result.

$$JAL_{macro} = \frac{\sum_{k=1}^{5} JAL_k}{N} \quad (5)$$

$$JAL_{micro} = 1 - \frac{\sum_{k=1}^{5} \sum_{i=1}^{n} p_i q_i}{\sum_{K}^{5} \sum_{i=1}^{n} p_i^2 + \sum_{K}^{5} \sum_{i=1}^{n} q_i^2} \quad (6)$$

where $k$ represents the classes of skin lesion attributes. This loss is effective to deal with multi-classes segmentation task with imbalance dataset.

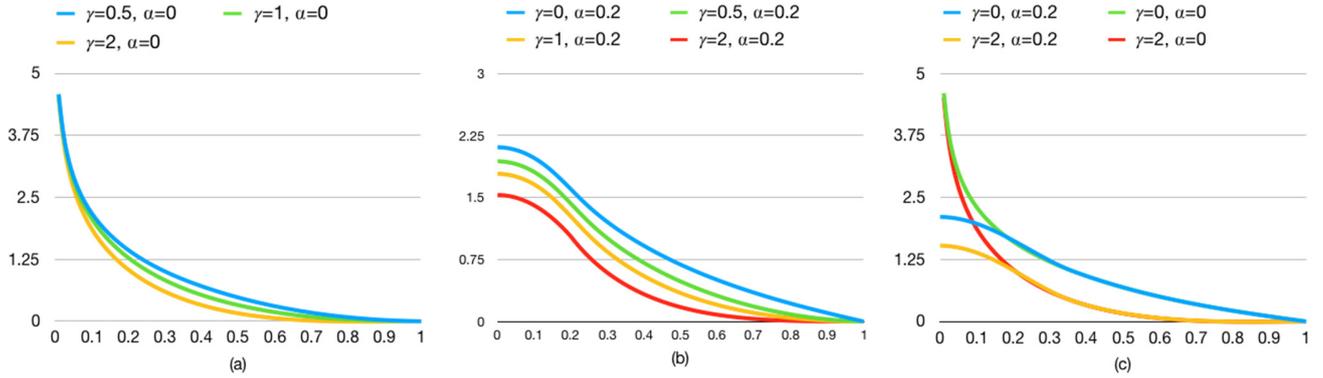

Figure 3. (a) the value of loss, when $\alpha=0$, $\gamma=0.5,1,2$; (b) the value of loss, when $\alpha=0.2$, $\gamma=0,0.5,1,2$; (c) the value of loss, when $\alpha=0$, $\gamma=0, 2$ and $\alpha=0.2$, $\gamma=0, 2$.

**Global balancing cross entropy loss**

The existence of irregular objects and inconsistent resolution renders images inevitably to contain indistinguishable parts, which leads to subjective annotations such as mislabeled objects (Zhou et al. 2019). Additionally, in order to enhance the accuracy of detecting skin lesion attributes, we incorporate the superpixel regional features by pooling feature maps according to superpixel maps, nevertheless, it has the risk of introducing noises. Accordingly, the superpixel-wise skin lesion attributes inevitably contain imperfect labels, which is harmful to deep convolutional networ training for the two following aspects. Firstly, the gradients of these inaccurate labels tend to over-whelm other informative labels in loss accumulation and lead the network to get stuck into local optimal solution. Secondly, the forcing learning of these subjective labels force network to learn non-robust, even trivial features so that the overfitting will be caused. In addition, the distribution of the extremely imbalanced class may make network overwhelmed by the negative samples.

To deal with the problems, our aim is to reduce the interferences of outliers and negative samples by modulating their contributions during the process of loss accumulation. Inspired by focal loss and smooth truncated loss, we propose a global balancing loss for skin lesion attribute detection. Global balancing loss endows the network with the ability to robustly detect features by focusing on learning informative samples. Therefore, the network can obtain better generalization capability.

$$GBL = \begin{cases} -\mathcal{F}(\alpha) + \frac{1}{2}\left(1 - \frac{p_t^2}{\alpha^2}\right), & p_t < \theta \\ -\mathcal{F}(p_t), & \theta < p_t \end{cases} \quad (7)$$

$$\mathcal{F}(p_t) = -(1-p_t)^\gamma \log(p_t) \quad (8)$$

where $p_t$ indicates the predicted probability, $p_t = p$ if $y = 1$ and $p_t = 1 - p$ if $y = 0$. $\theta$ is a boundary which divides global balancing loss into two parts. $\mathcal{F}$ is the focal loss and $\omega$ is a class weight for balancing each lesion attribute. As can be seen in Figure 3, global balancing loss harmonizes outliers when $p_t < \alpha$ and avoid network overwhelmed with negative samples. In this way, our network can focus on learned and informative samples. Therefore, our network could obtain better generalization than those under traditional cross entropy.

After defining the global balancing loss, we can finally have the overall target function:

$$L_{all} = 0.5 * GBCEL + 0.5 * GBJAL \quad (9)$$

where GBL represents global balancing loss, and JAL is defined as:

$$JAL = 1 - \frac{\sum_{i=1}^n p_i q_i}{\sum_{i=1}^n p_i^2 + \sum_{i=1}^n q_i^2} \quad (10)$$

## Experiments and Results

**Dataset**

Our framework is conducted on the 2018 ISIC skin lesion challenge-lesion attribute detections[1]. There is a total of 2594 samples for training and 1000 images for testing. And the skin lesions contain five type of attributes: Globules, Milia like cyst, Negative network, Pigment network and Streaks.

**Experiment setting**

Our experiments are implemented on Pytorch frame with an NVIDIA TITAN X GPU and it takes about 10 hours to train the proposed model. The ADSGrad, a variant of Adam optimizer (mini-batch size = 5, weight decay = 0.0001, and momentum = 0.9) is used to optimize the target function. The total epoch is 100, the learning rate is 0.0001 and power decay schedule is set to 0.9.

---

1  https://challenge2018.isic-archive.com/live-leaderboards/

| Attribute type | The percentage of blank images | Pixel Ratio (foreground / total pixels) |
|---|---|---|
| Pigment network | 41 | 13.70 |
| Negative network | 93 | 0.67 |
| Milia like cyst | 74 | 1.27 |
| Globules | 77 | 3.09 |
| Streaks | 96 | 0.42 |

Table 1. Information of each type of attribute.

| method | Avg JA | Avg Dice |
|---|---|---|
| CE+JAL | 26.1 | 39.5 |
| GBL $\gamma = 0$ | 28.3 | 42.2 |
| GBL $\gamma = 0.5$ | 28.5 | 42.6 |
| GBL $\gamma = 1$ | 29.3 | 43.7 |
| GBL $\gamma = 2$ | 28.8 | 43.1 |

Table 2. CE+JAL vs global balancing loss with varying $\gamma$.

**Data augmentation**

To eliminate the effect of the small dataset, such as overfitting, imbalanced class distribution and subjective labeling, an extensive set of augmentations is incorporated. The lists of specially employed augmentations include the adding Gaussian noises from different random distributions and types, randomly contrast adjustment and random sharpness adjustment (blurring using the Gaussian filter of different sizes and sharpening by employing CLAHE and embossment).

**Metrics**

According to the standard of ISIC 2018 skin lesion challenge, we employ the average jaccard index (avg JA) as goal metric and the average dice index (avg DICE) is used as auxiliary metric. Note that the avg JA is the average of five attributes JA and micro average JA, where Micro-average JA aggregates the contributions of all attributes to compute the JA.

**Results**

We investigate a serial of methods to demonstrate the effectiveness of random shuffle module, superpixel average pooling, SAM and global average pooling.

**Global balancing cross entropy loss**

To focus on global balancing cross entropy loss, experiments are all based on ResUnet. The baseline adopts cross entropy with jaccard loss as main loss for optimization. It has been shown that global balancing loss with $\gamma = 1$ achieves the best result.

**Superpixel attention network**

The introduction of random shuffle module and SAMaids convolutional neural network in three aspects: (1) learning local discriminative visual, (2) alleviating the noisy features introduced by deconstructed images, (3) obtaining precise boundary. Noted, all experiments in Table 3 are treained with GBL.

| Network name | Avg JA | Avg DICE |
|---|---|---|
| UNet | 23.7 | 37.4 |
| DeepLab | 18.6 | 28.8 |
| DANet (Fu et al. 2018) | 19.1 | 29.6 |
| ResUNet | 29.3 | 43.7 |
| ResUNet+RSM | 34.3 | 48.7 |
| ResUNet+SAM | 36.2 | 51.8 |
| ResUNet+RSM+SAM | 37.4 | 53.1 |

Table 3. The results of different networks.

As shown in Figure 4, SANet is more precise. This is because SAM instead of learning trivial boundary information but focusing on the visual pattern of texture, which is a kind of robust features in attributes detection task. Also, it has been shown that RSM could force network learning visual discriminative features. Milia like cysts is a kind of attributes scatter in lesion area. Since they grow independently, network requires delicate features of local region. In this way, the segmentation result of the second image demonstrates that RSM could be useful not only in classification tasks, but also segmentation challenges. Noted the results of ResNet151, ResNetv2, DenseNet169, Ensemble are from (Koohbanani et al. 2018; Szegedy et al. 2017).

## Conclusion

In this work, we propose a SANet framework with global balancing loss to address challenges of extremely imbalanced labels and low inter-class and high intra-class issues. Our proposed method achieves the state- of -the-art performance on the ISIC 2018 Task 2 with single model. We believe our methods would be utilized in many medical imaging challenges.

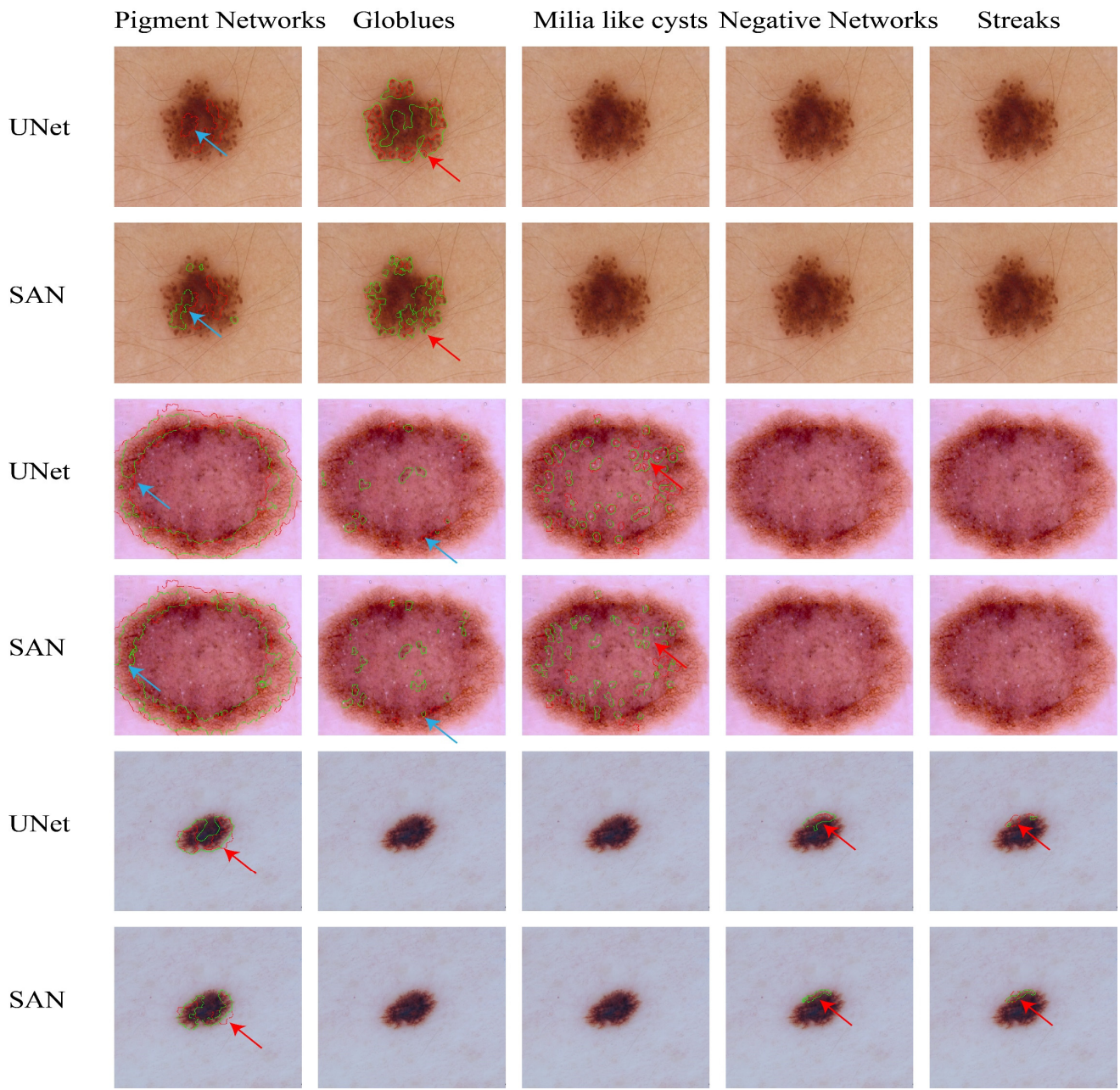

Figure 4. The improvements are achieved by SAN. The arrows highlight the difference between the prediction of UNet and SAN. The blue arrows indicate the benefits of RSM and red one point to SAM.

| Network name | Pigment Networks | | Globules | | Milia like Cysts | | Negative Networks | | Streaks | | Macro Avg | | Micro Avg | |
|---|---|---|---|---|---|---|---|---|---|---|---|---|---|---|
| | Jaccard | Dice | Jaccard | Dice | Jaccard | Dice | Jaccard | Dice | Jaccard | Dice | Jaccard | Dice | Jaccard | Dice |
| ResNet151 | 52.7 | 69.0 | 30.4 | 46.6 | 14.4 | 25.7 | 14.9 | 26.0 | 12.5 | 22.2 | 24.5 | 37.9 | | |
| ResNetv2 | 53.9 | 70.6 | 31.0 | 47.3 | 15.9 | 27.4 | 18.9 | 31.8 | 12.1 | 21.6 | 26.4 | 39.7 | | |
| DenseNet169 | 53.8 | 69.9 | 32.4 | 49.0 | 15.8 | 27.3 | 21.3 | 35.1 | 13.4 | 23.6 | 27.3 | 41.0 | | |
| Ensemble | 56.3 | 72.0 | 34.1 | 50.8 | 17.1 | 28.9 | 22.8 | 37.1 | 15.6 | 27.0 | 29.2 | 43.2 | | |
| SANet | 57.6 | 73.5 | 34.6 | 51.5 | 25.1 | 40.2 | 28.6 | 45.2 | 24.8 | 39.8 | 34.14 | 50.2 | 53.7 | 68.4 |

Table 4. Comparison of the results of proposed method and results presented in the literature.